\documentclass[pra, showkeys, reprint, nofootinbib]{revtex4-1}
\usepackage[english]{babel}
\usepackage[utf8]{inputenc}
\usepackage[colorinlistoftodos, color=green!40, prependcaption]{todonotes}
\usepackage{amsthm}
\usepackage{mathtools}
\usepackage{amsmath}
\usepackage{physics}
\usepackage{xcolor}
\usepackage{graphicx}
\usepackage{mathbbol}
\usepackage[pdftex, pdftitle={Article}, pdfauthor={Author}]{hyperref} % For hyperlinks in the PDF

\begin{document}
\title{Reachability Deficits in Quantum Approximate Optimization}

\author{V.~Akshay}
    \email[e-mail:]{akshay.vishwanathan@skoltech.ru}% Your name
      \homepage{http://quantum.skoltech.ru}
    \affiliation{Deep Quantum Laboratory, Skolkovo Institute of Science and Technology, 3 Nobel Street, Moscow, Russia 121205}
\author{H.~Philathong}
    \affiliation{Deep Quantum Laboratory, Skolkovo Institute of Science and Technology, 3 Nobel Street, Moscow, Russia 121205}
\author{M.E.S.~Morales}
    \affiliation{Deep Quantum Laboratory, Skolkovo Institute of Science and Technology, 3 Nobel Street, Moscow, Russia 121205}
\author{J.D.~Biamonte}
    \affiliation{Deep Quantum Laboratory, Skolkovo Institute of Science and Technology, 3 Nobel Street, Moscow, Russia 121205}

\date{\today} % Leave empty to omit a date

\begin{abstract}
The quantum approximate optimization algorithm (QAOA) has rapidly become a cornerstone of contemporary quantum algorithm development. Despite a growing range of applications, only a few results have been developed towards understanding the algorithms ultimate limitations. Here we report that QAOA exhibits a strong dependence on a problem instances constraint to variable ratio---this problem density places a limiting restriction on the algorithms capacity to minimize a corresponding objective function (and hence solve optimization problem instances). Such {\it reachability deficits} persist even in the absence of barren plateaus \cite{mcclean2018barren} and are outside of the recently reported level-1 QAOA limitations \cite{hastings2019classical}. These findings are among the first to determine strong limitations on variational quantum approximate optimization.
\end{abstract}

\maketitle

\paragraph*{Introduction.}\label{intro}
Variational hybrid quantum/classical algorithms have  become an area of significant interest \cite{Peruzzo2014}. These algorithms minimize objective functions which can be largely agnostic to systematic errors. This increases their potential in current Noisy Intermediate-Scale Quantum devices (NISQ) \cite{dicarlo2009demonstration,debnath2016demonstration,barends2016digitized}. These hybrid algorithms involve parameterized quantum circuits trained in a classical learning loop. Particular interest is the Quantum Approximate Optimization Algorithm (or QAOA) designed to find approximate solutions to combinatorial optimization problems \cite{farhi2014quantum}. Although QAOA has been shown to approximate solutions to problems such as MAX-CUT \cite{wang2018quantum} and realize Grover's search algorithm \cite{jiang2017,morales2018variational}, little is known about its ultimate limitations. 

Recent findings suggest that randomly parameterized quantum circuits in the large scale will suffer from \textit{barren plateaus} resulting in an exponentially low probability to find correct solutions \cite{mcclean2018barren}. Recent results also show that classical algorithms in a certain restrictive (single depth) setting can outperform level-1 QAOA \cite{hastings2019classical}. Can higher depth versions achieve advantage over classical algorithms?  

QAOA performance has an evident dependence on the circuit-depth and it is observed that increasing depth improves the quality of the possible approximation (at the cost of increasing the parameter search space). We show that circuit depth is not the only limiting restriction. Indeed, we found that finding appropriate solutions has strong dependence on the ratio of a problems constraint to variables (problem density). Hence, QAOA exhibits strong dependence on a problems density and for any fixed ansatz, there exists problem instances of high-density that appear not to be accessible. This feature persists as a fundamental limitation exhibited by QAOA.   

As a means to study the performance of QAOA, we turn to constraint satifiability-- a tool with a successful history.  Such problems are expressed in terms of \textit{n} variables and \textit{m} clauses (or constraints). The density of such problem instances is the clause to variable ratio, the clause density $\alpha = m/n$. $k$-SAT clauses are randomly generated to form instances by uniformly selecting unique $k$-tupels from the union of a variable set (cardinality $n>k$) and its element wise negation.  We consider both random instances of the NP-complete decision problem 3-SAT, as well as random instances of 2-SAT which is efficiently solvable. QAOA aims to approximate solutions to optimization version of these problems, MAX-2-SAT and MAX-3-SAT. Both are NP-Hard for exact solutions and APX-complete for approximations beyond a certain ratio \cite{haastad2001some}. In these settings, the algorithm's limiting performance exhibits strong dependence on the problem density in both cases.  

We call the QAOA problem general when considering random 3-SAT and 2-SAT instances with the standard one-body driver Hamiltonian \cite{farhi2014quantum}. In both cases we found strong limiting dependence of QAOA for clause densities above $\sim 1$. Moreover, the difference between the two problems (i.e.~2- vs 3-SAT) seemed negligible to a visual approximation. The density played the dominate role correlating inversely with performance. We further consider this same scenario, replacing the driver Hamiltonian with a $n$-body projector $(\ketbra{+}{+})^{\otimes{n}}$. While problem density dependence is still strongly exhibited, we found a decrease in the error of best-possible-approximation. Finally, by considering a single projector onto a solution space and the same driver as above, the variational version of Grover's search algorithm is recovered. While the clause density is fixed for a given $n$, the analytical solutions of this model provide a test bed to ascertain that energy approximation is critically dependent on circuit depth at each fixed density.  

\newpage

\paragraph*{Quantum Approximate Optimization.}\label{qaoa}
The usual procedure in implementing QAOA is as follows \cite{farhi2014quantum}:
\begin{enumerate}
    \item Create ansatz states, $\ket{\psi(\boldsymbol{\gamma},\boldsymbol{\beta})}$ on a quantum computer where, $\boldsymbol{\gamma}=(\gamma_{1},\gamma_{2},....,\gamma_{p})$ and $\boldsymbol{\beta}=(\beta_{1},\beta_{2},....,\beta_{p})$ are tunable over some fixed range. The state is prepared by applying a sequence of 2\textit{p}-parameter gates acting on the reference state, $\ket{+}^{\otimes{n}}$ as, 
    \begin{equation}\label{qaoa equation}
    \ket{\psi(\boldsymbol{\gamma},\boldsymbol{\beta})}=\prod_{i=1}^{p} \mathcal{U}(\gamma_{i},\beta_{i})\ket{+}^{\otimes{n}},
    \end{equation} 
    where
    \begin{equation}\label{driverandproblem}
        \mathcal{U}(\gamma_{k},\beta_{k})=\exp{-\imath \beta_{k}\mathcal{H}_{x}} \cdot \exp{-\imath \gamma_{k}\mathcal{V}}.
    \end{equation}
    \item Measurement of this state \eqref{qaoa equation} is done to compute the expected value of the objective function of interest, $\langle \mathcal{V}\rangle$. For QAOA, the objective function is the optimization  problem.
    \item Classical optimization algorithms are used to assign a set of parameters, $\boldsymbol{\gamma}^{*}$ and $\boldsymbol{\beta}^{*}$ that minimize $\bra{\psi(\boldsymbol{\gamma},\boldsymbol{\beta})}\mathcal{V}\ket{\psi(\boldsymbol{\gamma},\boldsymbol{\beta})}$. 
    \item Steps 1 and 2 are repeated by adjusting parameters to approximately minimize $\mathcal{V}$,   $\bra{\psi(\boldsymbol{\gamma}^{*},\boldsymbol{\beta}^{*})}\mathcal{V}\ket{\psi(\boldsymbol{\gamma}^{*},\boldsymbol{\beta}^{*})} \approx \min(\mathcal{V})$ . 
\end{enumerate}

The variational ansatz states created by QAOA take inspiration from the quantum adiabatic algorithm, where a system is initialized in an easy to prepare ground state of a local Hamiltonian $\mathcal{H}_{x}= \sum_{i} \sigma_{x}^{(i)},$ the driver Hamiltonian, which  is  then  slowly  transformed  to  the problem  Hamiltonian, $\mathcal{V}$ \cite{kadowaki1998quantum}. Trotterization of this procedure gives a long QAOA sequence as can be understood from equation \eqref{qaoa equation}. 
However, the Trotter approximation is evidently violated for a typical sequence. Outside of this understanding, the performance of QAOA seems rather remarkable.

\paragraph*{Quantum Approximation in Boolean Satisfiability.}\label{sat}
Boolean satisfiability is the problem of determining satisfiability of a Boolean expression written in conjunctive normal form (CNF). It is possible to map  any Boolean satisfiability problem into $3$-SAT; conjunction of clauses restricted to $3$ literals, via Karp reduction. It is well known that decision $3$-SAT is NP-complete  \cite{cook1971complexity}. Decision $2$-SAT, is the problem restricted to clauses limited to $2$ literals. This problem can be solved in polynomial-time \cite{krom1967decision}. $2$-SAT exhibits an algorithmic phase transition at a critical clause to variable ratio, $\alpha_{c} = 1$ \cite{goerdt1996threshold}. The transition is emperically exhibited in $3$-SAT numerics \cite{crawford1993experimental}. This implies that for $\alpha < \alpha_{c}$ almost all instances are satisfiable and for $\alpha  > \alpha_{c}$, almost all instances are not.

Known algorithms exhibit a slow-down around the phase transition, suggesting that most of the hard-instances are concentrated near this point. This signature of an easy-hard-easy transition is exhibited for the decision version of SAT. 

 In order to approximate solutions of MAX- $3$-SAT and MAX-$2$-SAT, an embedding scheme maps SAT-instances into Hamiltonians \cite{lucas2014ising,whitfield2012ground,biamonte2008nonperturbative} as : 

\begin{equation}\label{3sathamiltonian}
    \mathcal{H}_{\text{SAT}} = \sum_{l}\mathcal{P}(l),
\end{equation}

\noindent where \textit{l} indexes each clause in the SAT instance and $\mathcal{P}(l)$ are rank-one projectors that penalize each unsatisfiable assignments with at-least $1$ unit of energy. By this construction, we embed solutions to the MAX-$3$-SAT or MAX-$2$-SAT into the ground state space of $\mathcal{H}_{\text{SAT}}$. $2$-SAT instances requires only quadratic interactions whereas $3$-SAT requires $3$-body ones. Satisfiable instances are characterized by a zero ground state energy, $E_{g} = 0$ and unsatisfiable instances with $E_{g} \geq 1$, representative of the minimum violated clauses.

QAOA with standard settings, $\mathcal{H}_x=\sum_{i}\sigma_{x}^{(i)}$ and $\mathcal{V} = \mathcal{H}_{\text{SAT}}$, can now be used to calculate the energy approximation $E_{g}^{\text{QAOA}}$, where

\begin{equation}\label{qaoaoptimization}
    E_{g}^{\text{QAOA}} =  \min_{\boldsymbol{\gamma},\boldsymbol{\beta}} \bra{\psi(\boldsymbol{\gamma},\boldsymbol{\beta})}\mathcal{H}_{\text{SAT}}\ket{\psi(\boldsymbol{\gamma},\boldsymbol{\beta})}.
\end{equation}

\begin{figure*}[h!]

\minipage{\textwidth}
  \includegraphics[width=0.75\textwidth]{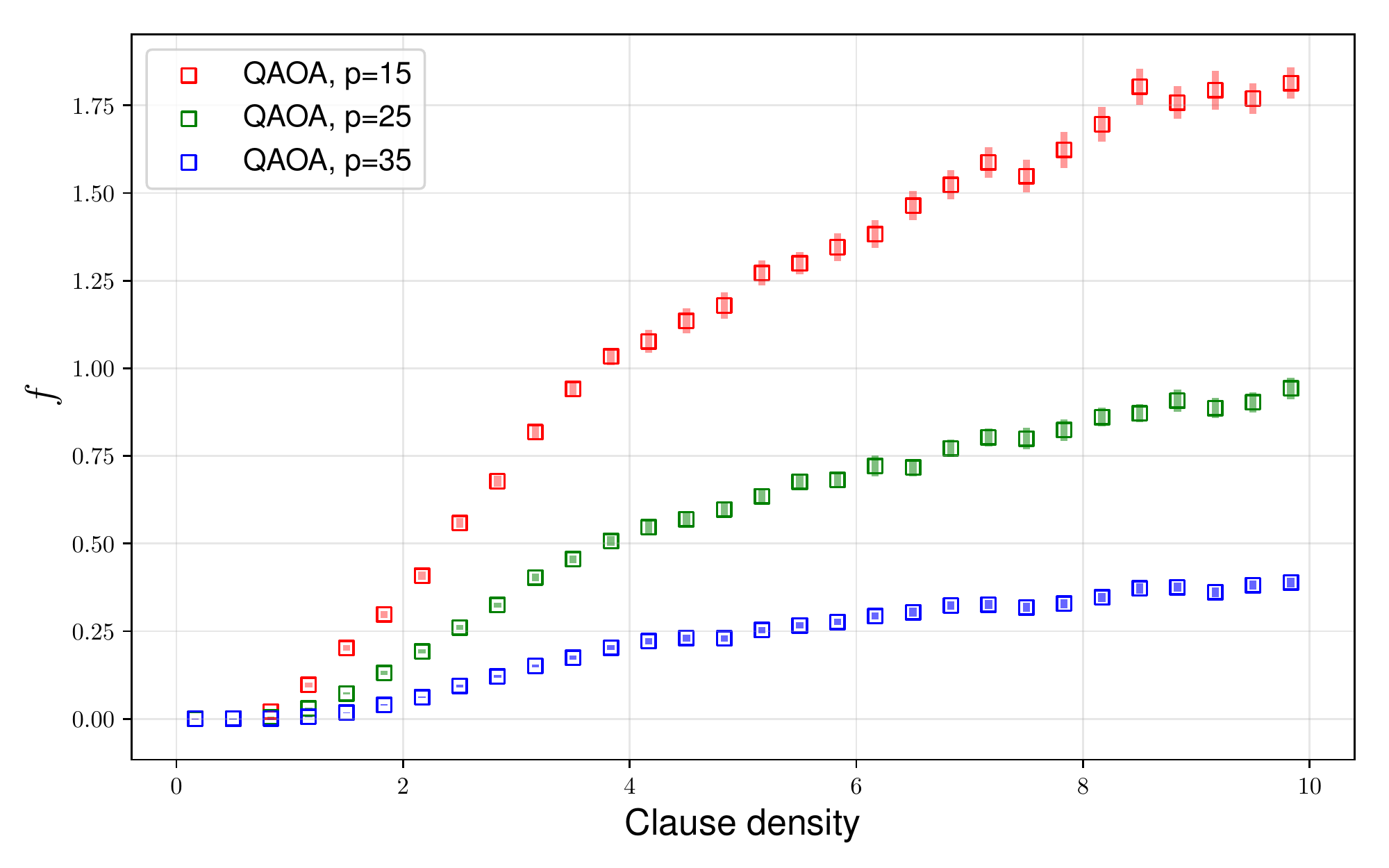}
  \includegraphics[width=0.75\textwidth]{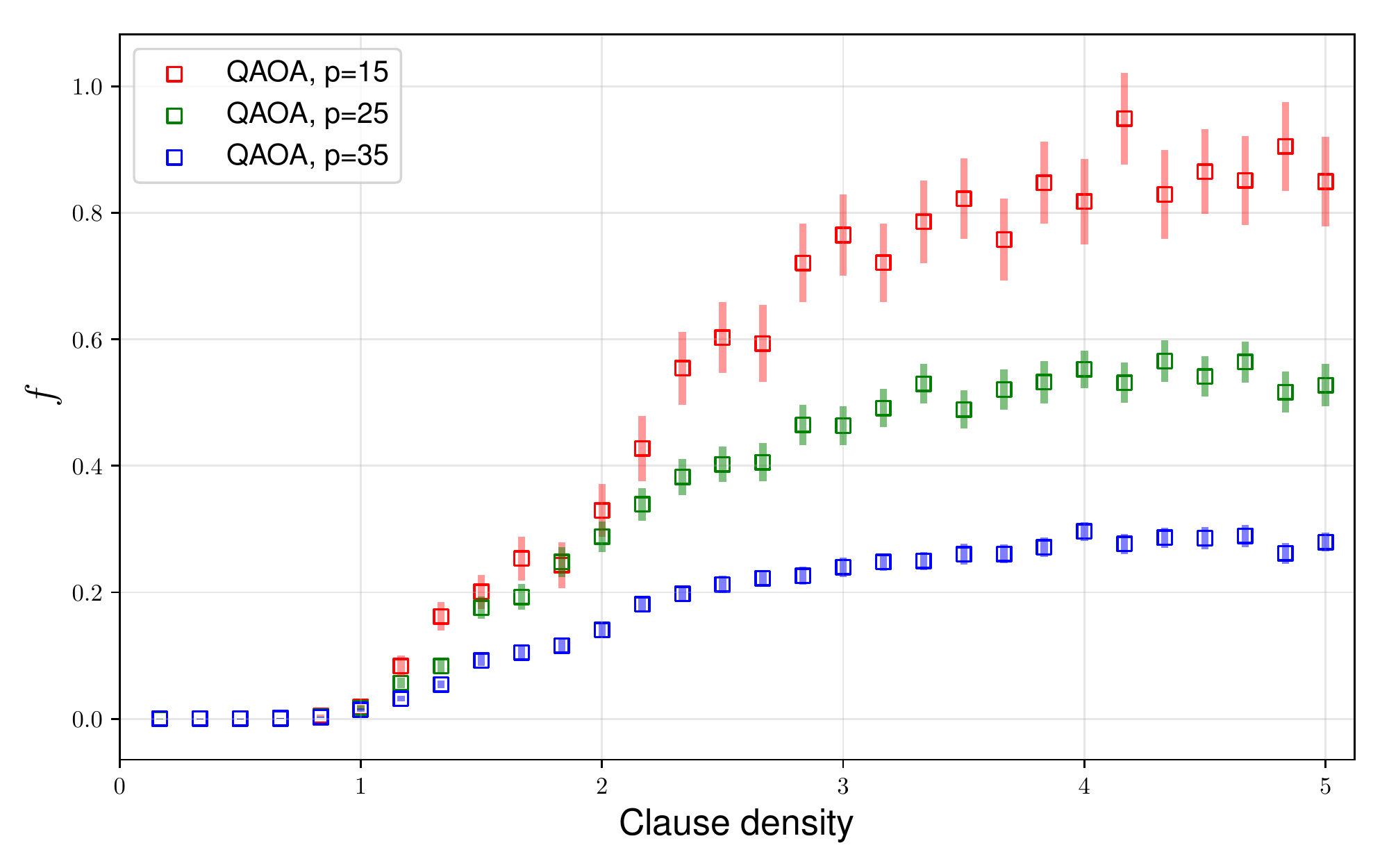}
\endminipage

\caption{$f = E_{g}^{\text{QAOA}} - \min(\mathcal{H}_{SAT})$ vs clause density for $3$-SAT (Top) and $2$-SAT (Bottom) for differing QAOA depths. Squares show the average value obtained over 100 randomly generated instances for $n=6$ with error bars indicating the standard error of mean. Plots also show improved performance for higher depths.}
\label{Standard QAOA}
\end{figure*}

 We numerically study $f = E_{g}^{\text{QAOA}} - \min(\mathcal{V}) $ as a function of clause density $\alpha$, for a \textit{p}-depth QAOA circuit on randomly generated $3$-SAT and $2$-SAT  instances (see Fig.~\ref{Standard QAOA}). Although increased depth versions achieve better approximations, the performance or best possible approximation of any arbitrary fixed depth QAOA exhibit a non trivial dependence on the density of the problem, $\alpha$.

\paragraph*{Reachability Deficits.}\label{rechabilitydeficits}

Fixed depth QAOA on MAX-$3$-SAT and MAX-$2$-SAT instances show limiting performance beyond a critical clause density. Higher depth versions are needed to break free from this limitation. Moreover, we observe recovery of ground state energy to strongly depend on the QAOA circuit depth and also the problem density for fixed problem size.  This limitation is what we refer to as \textit{reachability deficits} and is formulated as follows: \\

Let $\ket{\psi} $, be the ansatz states generated from a \textit{p}--depth QAOA circuit as shown in \eqref{qaoa equation}. Then 
\begin{equation}\label{reachabilitydef}
    f = \min_{\psi \subset \mathcal{H}} \bra{\psi}\mathcal{V}\ket{\psi} - \min_{\phi \in \mathcal{H}} \bra{\phi}\mathcal{V}\ket{\phi},
\end{equation}
characterises the limiting performance of QAOA. The R.H.S of equation \eqref{reachabilitydef} can be expressed as a function, $f(p,\alpha,n)$.
For $p \in \mathbb{N} $ and fixed problem size, $\exists$ $ \alpha > \alpha_c$ such that  $f \neq 0$. This is a reachability deficit. \\

\begin{figure*}[h!]

\minipage{\textwidth}

  \includegraphics[width=0.75\linewidth]{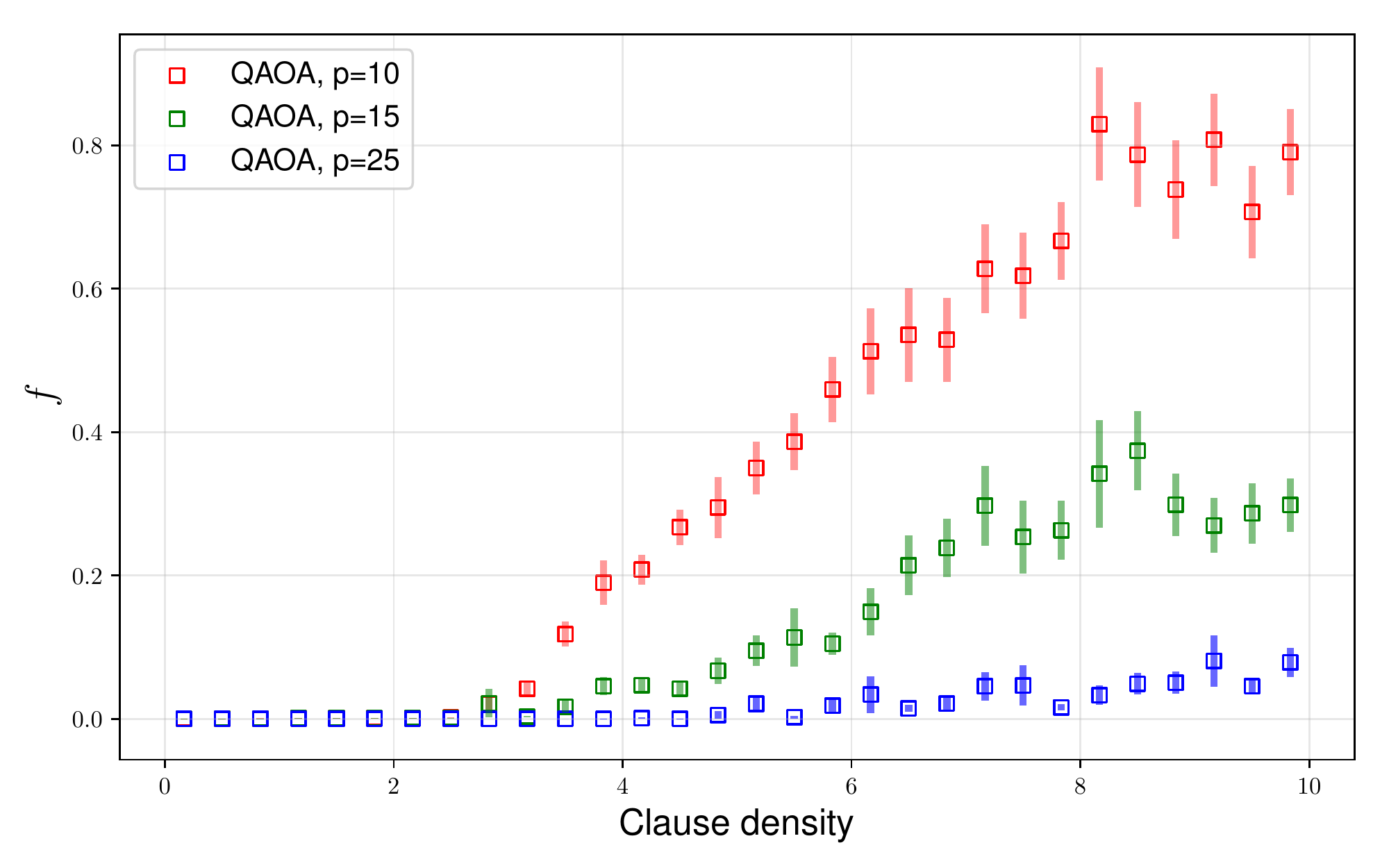} \\
  \includegraphics[width=0.75\linewidth]{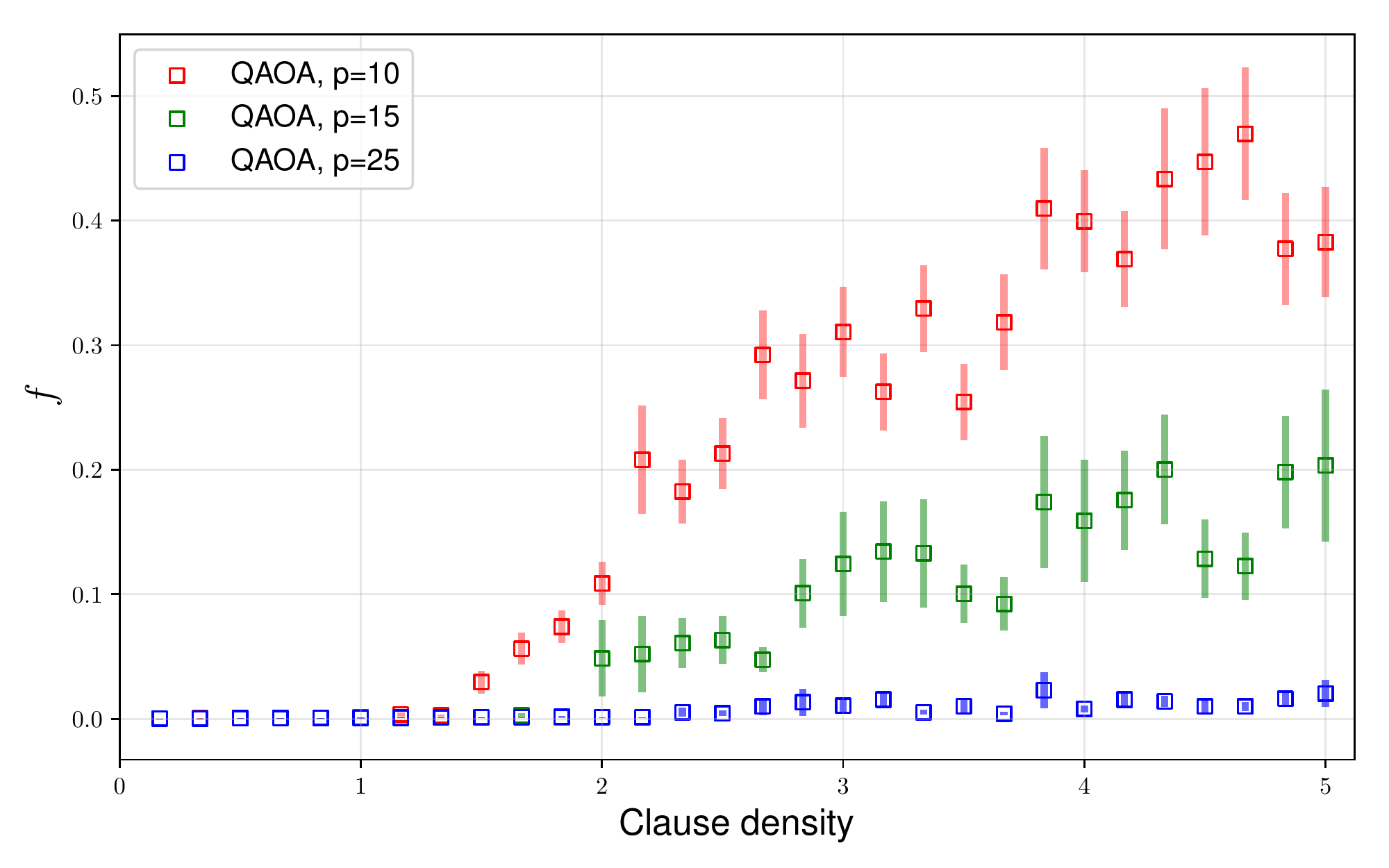}
\endminipage

\caption{$f = E_{g}^{\text{QAOA}} - \min(\mathcal{H}_{SAT})$ vs clause density for $3$-SAT (Top) and $2$-SAT (Bottom) for QAOA with driver Hamiltonian, $\mathcal{H}_{x} = (\ketbra{+}{+})^{\otimes{n}}$. Squares show the average value obtained over 100 randomly generated instances for $n=6$ with error bars indicating the standard error of mean. Plots also show improved performance for higher depths.}
\label{QAOA updated driver}

\end{figure*}

Fixed depth QAOA exhibit such reachability deficits even when modified with a new driver $\mathcal{H}_{x} = (\ketbra{+}{+})^{\otimes{n}}$ (see Fig.~\ref{QAOA updated driver}). The modified version requires lower circuit depths for achieving similar performance as standard QAOA but still exhibits reachability deficits.

\begin{figure*}[h!]

\minipage{\textwidth}

  \includegraphics[width=0.75\linewidth]{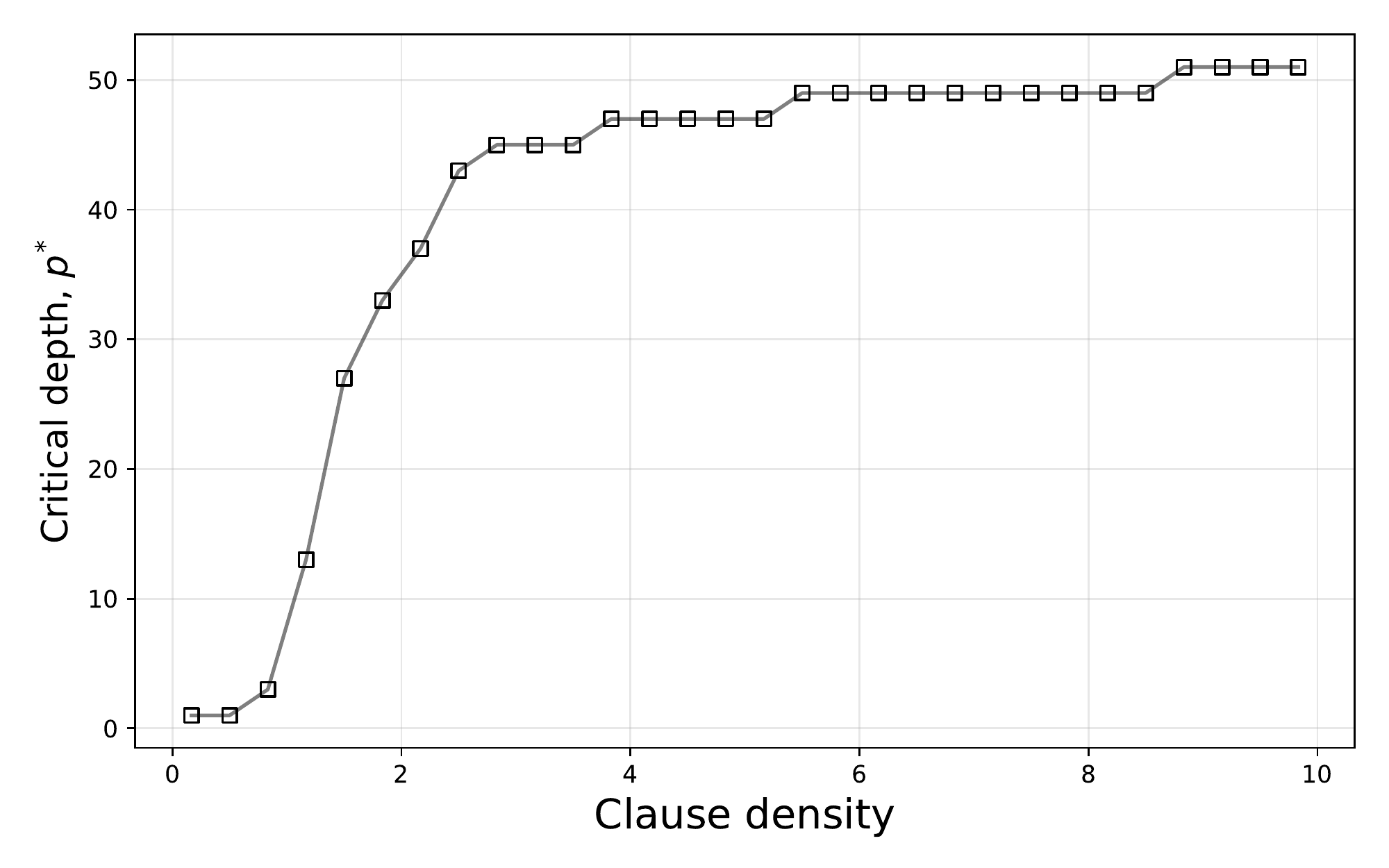} \\
  \includegraphics[width=0.75\linewidth]{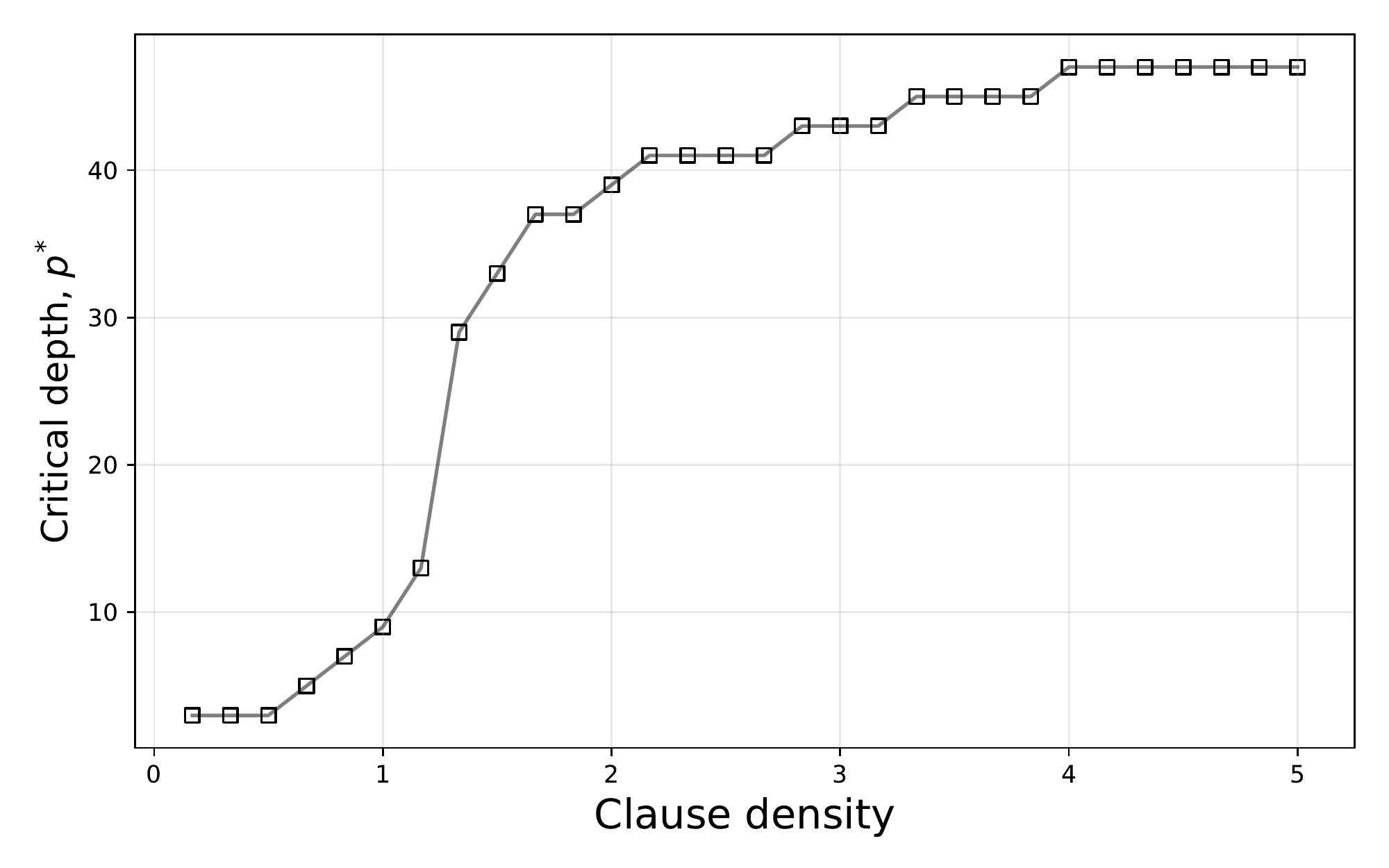}
\endminipage

\caption{Critical depth, $p^{*} $ vs clause density for $3$-SAT (Top) and $2$-SAT (Bottom). Squares represent the minimum QAOA depth for which an average overlap, $\eta \geq 0.95$ is achieved. Averages are calculated over 100 randomly generated SAT instances   }
\label{depthscale}

\end{figure*}

We analyse numerically the dependence of $p^{*}$, the critical circuit depth, or the minimum circuit depth for which QAOA returns the exact ground state energies (or least number of violated clauses) for both MAX-$3$-SAT and MAX-$2$-SAT up-to a set tolerance (see Fig.~\ref{depthscale}). The tolerance is set as a condition on the overlap between the QAOA generated state and the exact ground state which can be calculated as follows:\\
Let $\lbrace\ket{\text{gs}_{i}}\rbrace$ be the \textit{d} degenerate ground states of $\mathcal{H}_{\text{SAT}}$ then the overlap is given by,
\begin{equation}
    \eta = \sum_{i=1}^{d}{\abs{\bra{\psi_{p}(\boldsymbol{\gamma},\boldsymbol{\beta})}\ket{\text{gs}_{i}}}^{2}}.
\end{equation}
Instances to the left (classically easy),  $\alpha < \alpha_{c}$ require low critical depth to recover exact ground state energies and for instances, $\alpha > \alpha_{c}$, the critical depth is higher and grows with $\alpha$. Note in particular that the transition point does not coincide with the algorithmic phase transition point exhibited in decision version of 3-SAT, although the critical depth of exhibited reachability deficits is closer for 2-SAT's critical point.

Reachability deficits are different from barren plateaus \cite{mcclean2018barren}, where randomly parameterised quantum circuits for large problem sizes have exponentially low success in finding states that minimize the objective function. In the case of barren plateaus the state that achieves global minima of the objective function is accessible but choosing initial parameters randomly have greater probability to set the initial guess on a plateau of states where evaluation of gradients concentrates to zero. In contrast, irrespective of the initial parameter setting, QAOA with depth $p < p^{*}$ cannot reach optimal values as the corresponding state that achieves it becomes inaccessible.

\paragraph*{Variational Grover Search.}\label{toymodel}
Variational Grover search \cite{morales2018variational,grover1996fast} can be thought of as QAOA  with the following setting,
\begin{center}
$\mathcal{V}=\ketbra{\omega}{\omega}$ 
\end{center}
and 
\begin{center}
$\mathcal{H}_x=(\ketbra{+}{+})^{\otimes{n}}$,
\end{center}
where $\ket{\omega} \in \mathbb{C}_{2}^{\otimes{n}}$ is the objective state we are searching for. Hence the objective function of interest here is the minimization of the expected value of the Hamiltonian $\mathcal{H} = \mathbb{1} - \ketbra{\omega}{\omega}$ over the QAOA ansatz state $\ket{\psi(\boldsymbol{\gamma},\boldsymbol{\beta})}$ or,
\begin{equation}\label{objectivegrover}
     \min_{\boldsymbol{\gamma},\boldsymbol{\beta}}\langle \left( \mathbb{1} - \ketbra{\omega}{\omega} \right) \rangle_{\ket{\psi( \boldsymbol{\gamma} , \boldsymbol{\beta})}} = \min_{\boldsymbol{\gamma},\boldsymbol{\beta}} \left(1 - \abs{\bra{\omega}\ket{\psi(\boldsymbol{\gamma},\boldsymbol{\beta})}}^{2}\right). 
\end{equation}

The unitary gates that appear in equation \eqref{driverandproblem} can be simplified into the following expressions;
\begin{equation}
\begin{split}
\exp{-i\gamma_{k}\mathcal{V}}&= \exp{-i\gamma_{k}\ketbra{\omega}{\omega}}    \\
&=\mathbb{1} + \left(e^{-i\gamma_{k}} - 1 \right)\ketbra{\omega}{\omega},
\end{split}
\end{equation}
similarly,
\begin{equation}
\begin{split}
 \exp{-i\beta_{k}\mathcal{H}_{x}}&= \exp{-i\beta_{k}\ketbra{+}{+}^{\otimes{n}}}    \\
&=\mathbb{1} + \left(e^{-i\beta_{k}} - 1 \right)\ketbra{+}{+}^{\otimes{n}} .  
\end{split}
\end{equation}
We can then write the prepared ansatz state from a \textit{p}-depth QAOA circuit as, 
\begin{equation}\label{toymodel}
    \ket{\psi_{p}(\boldsymbol{\gamma},\boldsymbol{\beta})}=\text{A}_{p}\frac{1}{\sqrt{\text{N}-1}}\sum_{x \neq \omega}\ket{x} + \text{B}_{p}\ket{\omega},
\end{equation}
where, the amplitudes of one step can be related to the amplitudes of the next step via the recursive application of the matrix, 
\begin{equation}
    M_{p}=\quad
    \begin{pmatrix}
1 + \frac{a\left(\text{N}-1\right)}{N} & -a\left(b+1\right)\frac{\sqrt{\text{N}-1}}{N} \\
-a\frac{\sqrt{\text{N}-1}}{N} & \left(b+1\right)\left(1+\frac{a}{\text{N}}\right)
\end{pmatrix}.
\end{equation}
Here, $a=e^{-i\gamma_{p}}-1$,  $b=e^{-i\beta_{p}}-1$ and $\text{N}=2^{n}$. Substituting equation \eqref{toymodel} in equation \eqref{objectivegrover} we obtain the approximated energy as 
\begin{equation}
    E^{\text{QAOA}}_{g}= 1-\abs{B_{p}}^{2}.
\end{equation}

\begin{figure*}[h]

\minipage{\textwidth}
  \includegraphics[width=0.75\linewidth]{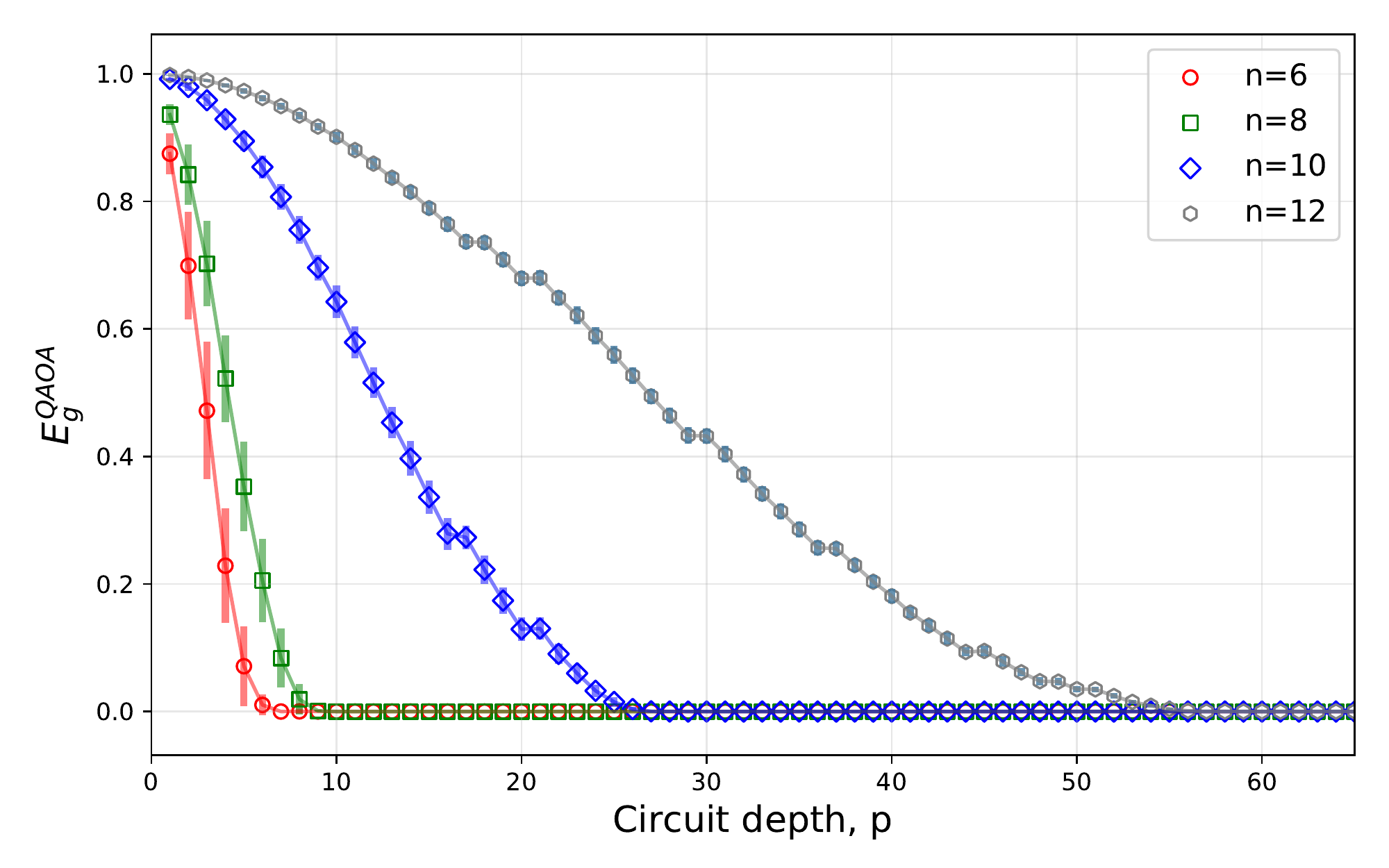}
\endminipage

\minipage{\textwidth}
  \includegraphics[width=0.7\linewidth]{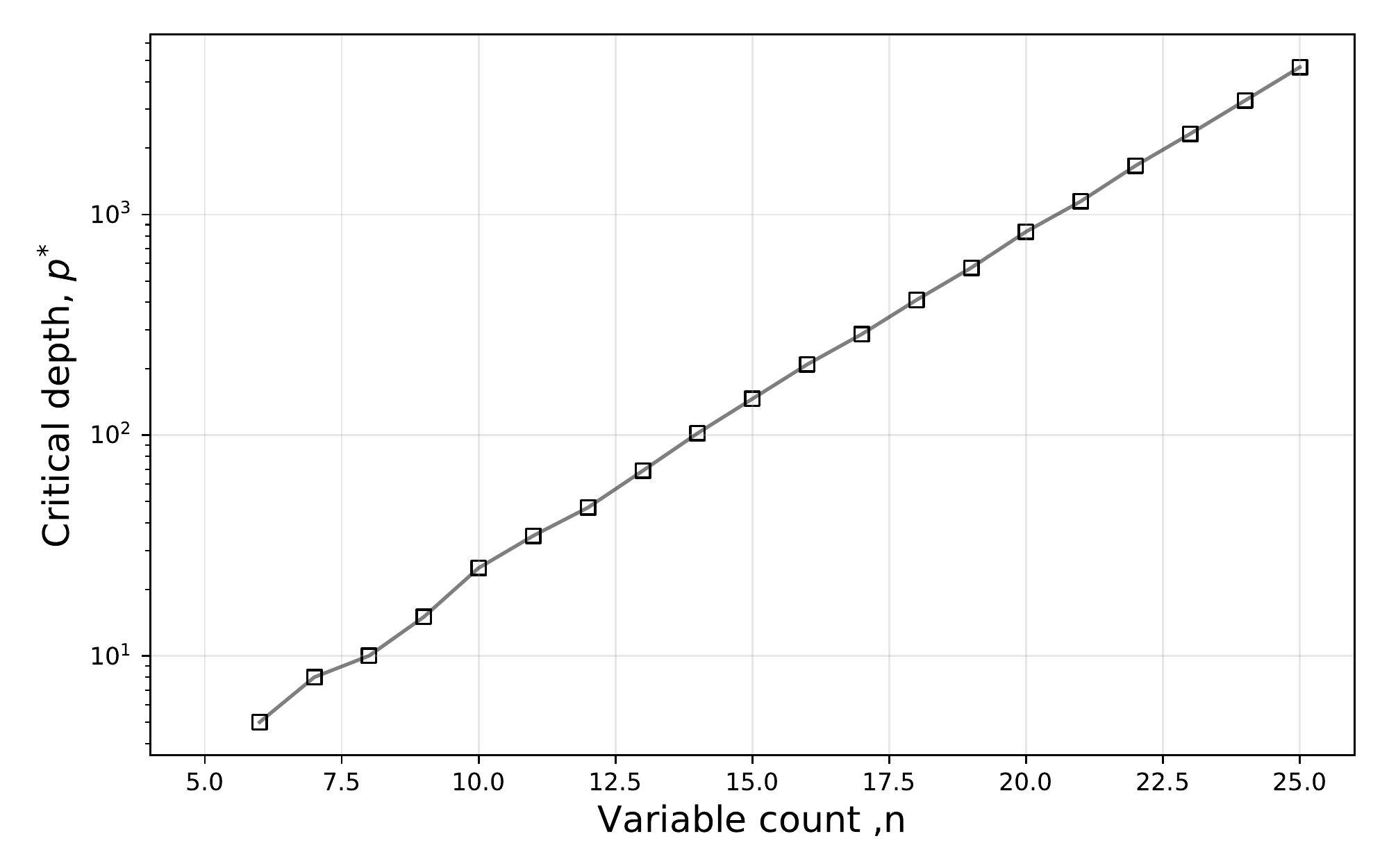}
\endminipage
\caption{Top: Convergence to exact ground state energy as a function of QAOA circuit depth for the variational Grover search on search space sizes, $n=6,8,10$ and $12$.
Bottom: Scaling of critical depth $p^{*}$ with variable count, $n$.}
\label{Variational Grover}
\end{figure*}

Minimization in equation \eqref{objectivegrover} is done numerically and the approximate energy as a function of circuit-depth is computed (see Fig.~\ref{Variational Grover}). For each problem size it is observed that approximated energy converges to the exact ground state energy $E_{g}=0$, when the circuit-depth reaches the critical value, $p^{*}$. At this depth, QAOA is able to exactly recover $\ket{\psi_{p^{*}}(\boldsymbol{\gamma^{*}},\boldsymbol{\beta^{*}})}=\ket{\omega}$ for some set of parameters $\boldsymbol{\gamma^{*}}$ and $\boldsymbol{\beta^{*}}$. 

If we set $p < p^{*}$, the minimization in equation \eqref{objectivegrover} terminates with $E_{g}^{QAOA} > 0$. This implies that in equation \eqref{toymodel}, $\abs{A_{p}}^{2} \neq 0$. It is evident that in such a case, QAOA cannot reach the state $\ket{\omega}$. The reachability deficit is removed only when the QAOA circuit is set with $p \geq p^{*}$. To establish the dependence of $p^{*}$ on the problem density, we increase \textit{n}, the size of the search space and recover a Grover scaling, $\mathcal{O}(\sqrt{N})$ for $p^{*}$.

\paragraph*{Discussion.}\label{discussions}
QAOA has been applied many times throughout the literature, with many findings reporting its surprising success.  However, findings to date appear to be implicitly constrained to instances of low-problem density (the ratio of an instances constraints to variables).  Hence, considered instances are not representative of statistical likely examples, and are only representative of the low-density subset. It is precisely this low-density subset which appears not to exhibit reachability deficits.  

Although the decision version of 2-SAT differs considerably from 3-SAT (complexity classes P and NP-Complete respectively), their optimization versions, MAX-2-SAT and MAX-3-SAT (both NP-Hard) show a remarkable similarity in critical QAOA depth requirement (see Fig.~\ref{depthscale}).   

We observe that instances with relatively low clause density require low depth QAOA circuits when compared to high density instances to approximate the minimum of an objective function up-to a given accuracy. Treating the critical QAOA depth as the computational cost, this crossover from low depth to high depth circuits is similar to the behaviour of computational resources reported for classical MAX-SAT solvers \cite{zhang2001phase}. Indeed the exhibited point of this crossover in MAX-SAT \cite{coppersmith2004random} is rather different from the computational phase transition point in Boolean satisfiability, further investigations are necessary in order to report a MAX-SAT phase transition in QAOA.

\onecolumngrid

\appendix*

\newpage

\newpage

\twocolumngrid


\begin{thebibliography}{10}

\bibitem{mcclean2018barren}
Jarrod~R McClean, Sergio Boixo, Vadim~N Smelyanskiy, Ryan Babbush, and Hartmut
  Neven.
\newblock Barren plateaus in quantum neural network training landscapes.
\newblock {\em Nature communications}, 9(1):4812, 2018.

\bibitem{hastings2019classical}
Matthew~B Hastings.
\newblock Classical and quantum bounded depth approximation algorithms.
\newblock {\em arXiv preprint arXiv:1905.07047}, 2019.

\bibitem{Peruzzo2014}
Alberto Peruzzo, Jarrod McClean, Peter Shadbolt, Man-Hong Yung, Xiao-Qi Zhou,
  Peter~J. Love, Al{\'{a}}n Aspuru-Guzik, and Jeremy~L. O'Brien.
\newblock A variational eigenvalue solver on a photonic quantum processor.
\newblock {\em Nature Communications}, 5(1), July 2014.

\bibitem{dicarlo2009demonstration}
Leonardo DiCarlo, Jerry~M Chow, Jay~M Gambetta, Lev~S Bishop, Blake~R Johnson,
  DI~Schuster, J~Majer, Alexandre Blais, L~Frunzio, SM~Girvin, et~al.
\newblock Demonstration of two-qubit algorithms with a superconducting quantum
  processor.
\newblock {\em Nature}, 460(7252):240, 2009.

\bibitem{debnath2016demonstration}
Shantanu Debnath, Norbert~M Linke, Caroline Figgatt, Kevin~A Landsman, Kevin
  Wright, and Christopher Monroe.
\newblock Demonstration of a small programmable quantum computer with atomic
  qubits.
\newblock {\em Nature}, 536(7614):63, 2016.

\bibitem{barends2016digitized}
Rami Barends, Alireza Shabani, Lucas Lamata, Julian Kelly, Antonio Mezzacapo,
  Urtzi Las~Heras, Ryan Babbush, Austin~G Fowler, Brooks Campbell, Yu~Chen,
  et~al.
\newblock Digitized adiabatic quantum computing with a superconducting circuit.
\newblock {\em Nature}, 534(7606):222, 2016.

\bibitem{farhi2014quantum}
Edward Farhi, Jeffrey Goldstone, and Sam Gutmann.
\newblock A quantum approximate optimization algorithm.
\newblock {\em arXiv preprint arXiv:1411.4028}, 2014.

\bibitem{wang2018quantum}
Zhihui Wang, Stuart Hadfield, Zhang Jiang, and Eleanor~G Rieffel.
\newblock Quantum approximate optimization algorithm for maxcut: A fermionic
  view.
\newblock {\em Physical Review A}, 97(2):022304, 2018.

\bibitem{jiang2017}
Zhang Jiang, Eleanor~G. Rieffel, and Zhihui Wang.
\newblock Near-optimal quantum circuit for grover's unstructured search using a
  transverse field.
\newblock {\em Phys. Rev. A}, 95:062317, Jun 2017.

\bibitem{morales2018variational}
Mauro~ES Morales, Timur Tlyachev, and Jacob Biamonte.
\newblock Variational learning of grover's quantum search algorithm.
\newblock {\em Physical Review A}, 98(6):062333, 2018.

\bibitem{cook1971complexity}
Stephen~A Cook.
\newblock The complexity of theorem-proving procedures.
\newblock In {\em Proceedings of the third annual ACM symposium on Theory of
  computing}, pages 151--158. ACM, 1971.

\bibitem{kadowaki1998quantum}
Tadashi Kadowaki and Hidetoshi Nishimori.
\newblock Quantum annealing in the transverse {Ising} model.
\newblock {\em Physical Review E}, 58(5):5355, 1998.

\bibitem{krom1967decision}
Melven~R Krom.
\newblock The decision problem for a class of first-order formulas in which all
  disjunctions are binary.
\newblock {\em Mathematical Logic Quarterly}, 13(1-2):15--20, 1967.

\bibitem{goerdt1996threshold}
Andreas Goerdt.
\newblock A threshold for unsatisfiability.
\newblock {\em Journal of Computer and System Sciences}, 53(3):469--486, 1996.

\bibitem{crawford1993experimental}
James~M Crawford and Larry~D Auton.
\newblock Experimental results on the crossover point in satisfiability
  problems.
\newblock In {\em Artificial Intelligence}, volume~81, pages 31--57, 1996.

\bibitem{lucas2014ising}
Andrew Lucas.
\newblock Ising formulations of many NP problems.
\newblock {\em Frontiers in Physics}, 2:5, 2014.

\bibitem{whitfield2012ground}
James~Daniel Whitfield, Mauro Faccin, and JD~Biamonte.
\newblock Ground-state spin logic.
\newblock {\em EPL (Europhysics Letters)}, 99(5):57004, 2012.

\bibitem{biamonte2008nonperturbative}
JD~Biamonte.
\newblock Nonperturbative k-body to two-body commuting conversion
  {Hamiltonians} and embedding problem instances into {Ising} spins.
\newblock {\em Physical Review A}, 77(5):052331, 2008.

\bibitem{grover1996fast}
Lov~K Grover.
\newblock A fast quantum mechanical algorithm for database search.
\newblock {\em arXiv preprint quant-ph/9605043}, 1996.



\bibitem{haastad2001some}
Johan H{\aa}stad.
\newblock Some optimal inapproximability results.
\newblock {\em Journal of the ACM (JACM)}, 48(4):798--859, 2001.

\bibitem{zhang2001phase}
Weixiong Zhang.
\newblock Phase transitions and backbones of 3-sat and maximum 3-sat.
\newblock In {\em International Conference on Principles and Practice of
  Constraint Programming}, pages 153--167. Springer, 2001.

\bibitem{coppersmith2004random}
Don Coppersmith, David Gamarnik, MohammadTaghi Hajiaghayi, and Gregory~B
  Sorkin.
\newblock Random max sat, random max cut, and their phase transitions.
\newblock {\em Random Structures \& Algorithms}, 24(4):502--545, 2004.


\end{thebibliography}
\end{document}